\documentstyle[12pt]{article}
\newcommand{\be}{\begin{equation}}
\newcommand{\ee}{\end{equation}}
\newcommand{\bea}{\begin{eqnarray}}
\newcommand{\eea}{\end{eqnarray}}
\newcommand{\ba}{\begin{array}}
\newcommand{\ea}{\end{array}}

\def\bbox{{\,\lower0.9pt\vbox{\hrule \hbox{\vrule height 0.2 cm
\hskip 0.2 cm \vrule height 0.2 cm}\hrule}\,}}
\newcommand{\dsl}{\pa \kern-0.5em /}

\font\mybb=msbm10 at 12pt
\def\bb#1{\hbox{\mybb#1}}

\def\bN {\bb{N}}
\def\bP {\bb{P}}
\def\bH {\bb{H}}

\def\bL {\bb{L}}

\def\bbQ {\bb{Q}}

\def\appendix#1{
  \addtocounter{section}{1}
  \setcounter{equation}{0}
  \renewcommand{\thesection}{\Alph{section}}
  \section*{Appendix \thesection\protect\indent \parbox[t]{11.15cm}
  {#1} }
  \addcontentsline{toc}{section}{Appendix \thesection\ \ \ #1}
  }

\begin{document}



\begin{titlepage}
\rightline{DAMTP-2007-89}
\rightline{\tt{arXiv:0710.5178}}

\vfill

\begin{center}
\baselineskip=16pt 
{\Large{\bf Hamilton-Jacobi  Mechanics from}}

\vskip 0.2cm

{\Large{\bf Pseudo-Supersymmetry}}

\vskip 0.3cm
{\large {\sl }}
\vskip 10.mm
{\bf ~ Paul K. Townsend}
\vskip 1cm
{\small
Department of Applied Mathematics and Theoretical Physics\\
Centre for Mathematical Sciences, University of Cambridge\\
Wilberforce Road, Cambridge, CB3 0WA, UK\\
}
\end{center}
\vfill

\par
\begin{center}
{\bf ABSTRACT}
\end{center}
\begin{quote}

For a general mechanical system, it is shown that each solution of the Hamilton-Jacobi equation defines 
an $N=2$ pseudo-supersymmetric extension of the system, such that the usual relation of the momenta to Hamilton's principal function is  the `BPS' condition for  preservation of 1/2 pseudo-supersymmetry.  The examples of the relativistic  and non-relativistic particle, in a general potential, are worked through in detail and used to discuss the relation to cosmology and to supersymmetric quantum mechanics.

\vfill
\end{quote}
\end{titlepage}

\section{Introduction}
\setcounter{equation}{0}

The reparametrization invariant  dynamics of a Hamiltonian  system with $s$ degrees of freedom is determined by a Hamiltonian constraint on a $(2s+2)$-dimensional phase space. Given
local Darboux coordinates $(X^m;P_m)$ ($m=0,1,\dots,s$), the Lagrangian 
takes the form
\be\label{startlag}
L= \dot X^m P_m  - \ell \, {\cal H}\left(X; P\right)\, , 
\ee
where the overdot indicates differentiation with respect to an arbitrary  time parameter, and 
$\ell$ is a Lagrange multiplier for the Hamiltonian constraint.  The Hamilton equations of motion 
are then
\be\label{HEM}
\ell^{-1} \dot X^m = \frac{\partial {\cal H}}{\partial P_m} \, , \qquad
\ell^{-1} \dot P_m = - \frac{\partial {\cal H}}{\partial X^m}\, . 
\ee
The constraint function ${\cal H}$  will be assumed to be a  polynomial in the momenta $P$ that is at most quadratic. This is the case of most interest and most others can be put in this form by increasing the dimension of the phase space. Thus, 
\be\label{genH}
{\cal H} = \frac{1}{2} K^{mn}(X) P_mP_n + J^m(X)P_m + U(X)\, ,  
\ee
for symmetric tensor field $K(X)$, vector field $J(X)$ and scalar potential function $U(X)$. If $K$ is invertible it may be interpreted as the metric on a target space with local coordinates $X$. The case in which $K$ is not invertible\footnote{The possibility that $K(X)$ is invertible for some $X$ and non-invertible for other $X$ or, more generally, that the dimension of its kernel is position dependent, will not be considered here because neither it is considered  in expositions of  Hamilton-Jacobi theory.} is also of interest because this allows for some components of $P$ to appear  linearly, but then  we must insist that $v\cdot J\ne0$ for any co-vector field $v(X)$ in the kernel of $K$; in other word
\be\label{restrictJ}
v_mJ^m \ne 0 \qquad {\rm if} \qquad K^{mn}v_n =0\, . 
\ee
Otherwise, there are components of $P$ that do not appear in ${\cal H}$, so their conjugate variables are constants and we may  rewrite the model in terms of a lower-dimensional phase space.

The action on the constraint surface, viewed as a function of $X$ at a `final' time, is Hamilton's `principal' function:
\be
S(X)= \int^X \!dX^mP_m\, , 
\ee
from which we we deduce that
\be\label{foeqs}
P_m = \partial_m S \qquad (m=0,1,\dots,s), 
\ee
where $\partial_m=\partial/\partial X^m$.  These equations typically yield first-order differential equations for $X$ when combined with the equation of motion for $P$, so we will sometimes refer to them as the `first-order' equations of HJ theory.  Using (\ref{foeqs})  in the Hamiltonian constraint, we find the Hamilton-Jacobi (HJ) equation
\be\label{genHJ}
\frac{1}{2}K^{mn} \partial_m S\, \partial_n S + J^m\partial_m S + U(X) =0\, . 
\ee
The solutions of this equation are in one-to-one correspondence with solutions of (\ref{HEM}). 
This is not obvious, and Hamilton (who found the equation) did not appreciate this point, which is due to Jacobi.  The main aim of this paper is to present a new derivation of HJ theory, with various novel features, one of which is that  it  makes this feature of the HJ formalism manifest.  The new derivation relies on a correspondence between solutions of the HJ equation for a given mechanical system and $N=2$ pseudo-supersymmetric extensions of it. 

The concept of pseudo-supersymmetry has  its origins in  supergravity  \cite{Pilch:1985aw,deWit:1987sn}, and can be viewed as a complex analytic continuation of `standard' supersymmetry.  Its relevance to the Hamilton-Jacobi approach to inflationary cosmology was noticed in  \cite{Skenderis:2006jq,Skenderis:2006rr}, where the `first-order' equations of  the HJ approach to cosmology \cite{Salopek:1990jq}  were recovered from an effective relativistic particle mechanics model, and interpreted as integrability conditions for the existence of  `pseudo-Killing' spinors\footnote{Because of the ``Domain-Wall/Cosmology Correspondence'' \cite{Cvetic:1994ya,Skenderis:2006jq}, there is an analogous story for domain walls and many of the observations summarized here were first made in that context. However, in this introduction we focus on cosmology for simplicity of presentation.}. This prompted several  explicit realizations of pseudo-supersymmetric cosmologies in a (`variant') supergravity context, where the pseudo-Killing spinors become `genuine' Killing spinors associated with the partial preservation of a  symmetry \cite{Bergshoeff:2007cg,Skenderis:2007sm,Vaula:2007jk}. In this sense, the equations (\ref{foeqs}) for these models  may be interpreted as  `BPS' conditions, but this interpretation is not {\it intrinsic} to the effective particle mechanics model because it involves consideration of spacetime spinors. However, the supergravity examples do suggest the possibility of an intrinsic BPS interpretation, via an explicit pseudo-supersymmetric extension of the effective particle mechanics model. This was the starting point for the work reported here. 

It turns out that the required pseudo-supersymmetric extension of mechanics is not difficult to construct. As will be explained, there is a close relationship to standard supersymmetric mechanics \cite{Witten:1981nf}, but this suggests a difficulty: supersymmetry implies restrictions on the possible potentials, and one therefore expects the same to  be true of pseudo-supersymmetry. In contrast,  the HJ formalism does not involve any such restriction. Here, the key observation is that the superpotentials required for pseudo-supersymmetry may be multi-valued functions with branch points  \cite{Skenderis:2006jq,Sonner:2007cp}. In the context of the non-relativistic particle, the relation of this observation  to HJ theory is obvious, and it leads directly to the general pseudo-supersymmetry formalism described here. We present this formalism from a classical perspective, initially,  because the intention is  to gain insight into standard results of classical mechanics, which we do by focusing on solutions of the  equations of motion for which all anti-commuting variables are zero. 

In the quantum theory, one cannot set the anti-commuting variables to zero because this is not consistent with their anti-commutation relations. They could be `integrated out' (in a path-integral formulation) but this still  implies some contribution to the wave-function. The standard semi-classical wavefunction has the form $\rho \exp iS$ for some variable modulus $\rho$ determined by quantum fluctuations, so one would expect the anti-commuting variables to contribute to $\rho$. This is indeed the case, and the net result is that $\rho$ is constant: the quantum fluctuations of the new anti-commuting variables cancel  the quantum fluctuations of the original model. In effect, the quantum  theory of the pseudo-supersymmetric mechanics model  is equivalent to classical mechanics of the original model! This is reminiscent of the reformulation of classical mechanics of  Gozzi et al.  \cite{Gozzi:1986ge,Gozzi:1989bf} but the details appear to be rather different; in part because of the central role of the Hamilton-Jacobi equation in the formalism presented here. 

\subsection{Particle Mechanics Examples}

As an aid to understanding the general formalism to follow, it may help to keep in mind the case of a particle  in a space of dimension $s$, which yields various possible mechanical systems with $s$ degrees of freedom.  We shall later consider the $s=1$ case for (i) a relativistic particle and (ii) a non-relativistic particle. When needed, we use the following notation for the components of the phase superspace coordinates: 
\bea
X^m &=& (t,x)\, , \qquad \ P_m = (-E,p)\, , \nonumber\\
\Lambda^m &=& (\psi,\lambda)\, , \qquad \bar\Lambda_m=(-\bar\psi,\bar\lambda). 
\eea
The momentum $P$ is  a co-vector  in a 2-dimensional Minkoski spacetime with 
signature $(-1,1)$, and
\be
P^2 \equiv \eta^{mn}P_mP_n = -E^2 + p^2\, . 
\ee
The two cases to be considered later are defined as follows:
\begin{itemize}
\item {\bf Relativistic particle}:
\be\label{HR}
{\cal H} = \frac{1}{2} P^2 + U(X)\, . 
\ee
This illustrates the case in which $K$ is non-degenerate and $J=0$. 
For $2U=m^2$ we have a free relativistic particle of mass $m$, but we will  consider a more complicated potential that is relevant to cosmology. 

\item {\bf Non-relativistic particle}:
\be\label{HNR}
{\cal H} = -E + \frac{1}{2}p^2 + U(t,x)\, .  
\ee
This illustrates the case in which $K$ is degenerate and $J\ne0$. For $t$-independent potential we will write $U=V(x)$. 
\end{itemize}

\section{The Formalism}
 
We begin by supposing that  $(X;P)$ phase space is the `body' of a  phase superspace with coordinates $(X,\Lambda;P,\bar\Lambda)$, where the anticommuting $(s+1)$-vector $\Lambda$ and $(s+1)$-covector $\bar\Lambda$ are canonically conjugate. We now allow for an extension of the Hamiltonian constraint function ${\cal H}$ on the $(X;P)$ `body' of the phase superspace to a function $\bH$ on the full phase superspace, and then consider a Lagrangian of the form
\be\label{newlag}
\bL= \dot X^m P_m +i\bar\Lambda_m\dot\Lambda^m - \ell \, \bH + i\chi \bbQ +i\bar\chi \bar {\bbQ}
\ee
where $(\chi,\bar\chi)$ is a pair of real anticommuting Lagrange multipliers for a pair of 
constraints with {\it real} anti-commuting constraint functions $(\bbQ,\bar {\bbQ})$. 
The Lagrangian itself is real because we adopt the convention that complex conjugation changes the order of anti-commuting quantities.  It follows from this Lagrangian  that the non-zero Poisson brackets of the dynamical variables are
\bea
\{X^m,P_n\}_{PB} &\equiv& - \{P_n,X^m\}_{PB}  = \delta^m_n  \, , \nonumber\\
\{\Lambda^m,\bar\Lambda_n\}_{PB} &\equiv& \{\bar\Lambda_n,\Lambda^m\}_{PB} = 
-i\delta^m_n\, , 
\eea
and one may use this result to compute the Poisson brackets of the constraint functions, which must be in involution for consistency. We will require that the only non-zero Poisson bracket of  constraint functions is
\be\label{PBrelation}
\{\bbQ,\bar{\bbQ}\}_{PB} = -2i \bH\, . 
\ee
If  $\bar{\bbQ}$ were the complex conjugate of $\bbQ$ then we would have  a standard 
$N=2$ supersymmetric extension of the model defined by (\ref{startlag}). Instead, both $\bbQ$ and $\bar{\bbQ}$ are real, so we have an $N=2$ pseudo-supersymmetric extension. 

Each of the constraint functions generates a {\it local} symmetry. In particular, the infinitesimal
pseudo-symmetry variations of any function $\Phi$ on the phase superspace are given by
\be
\delta_\epsilon\Phi = i\left\{\epsilon\bbQ,\Phi\right\}_{PB} \, , \qquad
\delta_{\bar\epsilon}\Phi = i\left\{\bar\epsilon\bar{\bbQ},\Phi\right\}_{PB}\, , 
\ee
where the real anticommuting parameters $(\epsilon,\bar\epsilon)$ are arbitrary functions of the arbitrary independent variable. 

\subsection{Principal function as Superpotential} 

We shall now {\it re-interpret the HJ equation (\ref{genHJ}) as an expression for $U$ in terms of a superpotential $S$}. We may then rewrite the Hamiltonian constraint function in the factorized form
\be\label{facH}
{\cal H} =\frac{1}{2} \left[K^{mn}\left(P_n + \partial_n S\right) + 
2J^m\right] \left(P_m-\partial_m S\right)\, . 
\ee
We now introduce a symmetric affine connexion $\Gamma$ on the coordinate space such that  both the tensor $K$ and the vector $J$ are covariantly constant:
\be
\partial_p K^{mn} = - 2\Gamma^{(m} {}_{pq} K^{n)q}\, , \qquad \partial_p J^m = -\Gamma^m{}_{pq}J^q\, . 
\ee
When $K$ is invertible, and interpreted as a metric, the first of these conditions implies that $\Gamma$ is the usual Levi-Civita  connexion, and the second condition is then a constraint on the allowed choices for $J$.
The curvature tensor constructed from $\Gamma$ is 
\be
R^m{}_{npq} = 2 \partial_{[q}\Gamma^m{}_{p]n} + 2\Gamma^m{}_{\ell[q}\Gamma^\ell{}_{p]n}\, ,   
\ee
and it satisfies the cyclic identity $R^m{}_{[npq]} \equiv 0$. 

Now define\footnote{This shift of $P$ `covariantizes' the $\bar\Lambda\dot\Lambda$ term in the Lagrangian, so all constraint functions become manifestly covariant when expressed in terms of $\bP$.}
\be
\bP_n = P_n + i\Gamma^m{}_{np}\Lambda^p\bar\Lambda_m\, , 
\ee
and consider the choice
\bea\label{Qs}
\bbQ &=& \left(\bP_m-\partial_m S\right)\Lambda^m \equiv
\left(P_m-\partial_m S\right)\Lambda^m\nonumber \\
\bar{\bbQ} &=& \left[K^{mn}\left(\bP_n+ \partial_n S\right) + 2J^m \right] \bar\Lambda_m \, . 
\eea
It  is obvious that $\{\bbQ,\bbQ\}_{PB}=0$, and a computation of $\{\bar{\bbQ},\bar{\bbQ}\}_{PB}$ shows that this is zero too provided that
\be
K^{p[r}K^{n|\ell|} R^{m]}{}_{q\ell p} =0\, ,
\ee
which is a consequence of the cyclic identity for invertible $K$. Finally, a computation of 
$\{\bbQ,\bar{\bbQ}\}_{PB}$ shows that (\ref{PBrelation}) holds with
\bea
\bH &=& \frac{1}{2} \left[K^{mn}\left(\bP_n + \partial_n S\right) + 2J^m\right]\left(\bP_m -\partial_m S\right) \nonumber\\
&& \ -i K^{mp}\left({\cal D}_p\partial_n S\right) \Lambda^n\bar\Lambda_m + 
\frac{1}{4}K^{n\ell}R^m{}_{\ell pq}\, \Lambda^p\Lambda^q\bar\Lambda_n\bar\Lambda_m\, . 
\eea
As required, $\bH \to {\cal H}$ when all anti-commuting variables are set to zero.

To summarize: for every solution $S$ of the HJ equation of some given mechanical model, we have an $N=2$ pseudo-supersymmetric extension of that model in which Hamilton's principal function $S$ is re-interpreted as a superpotential.  There is something odd about this result: there was no restriction on the initial choice of potential $U$ but we are now saying that it should be expressible in terms of a superpotential, so should this not restrict the potential in some way? The sharpest illustration of this `paradox' is provided by the non-relativistic particle constrained to move on the $x$ axis in a time-independent potential $V(x)$. We shall be studying this example 
in detail; to anticipate, $N=2$ pseudo-supersymmetry implies that $V$ is given in terms of a superpotential $W(x)$ by the formula
\be\label{Vnon-rel}
V= E_0 - \frac{1}{2}\left(W'\right)^2\, , 
\ee
where $E_0$ is the particle's energy. Clearly, there are many potentials $V$ that cannot be written in this form; for example, the harmonic oscillator potential $V=x^2$. However, the potential $V$  is only constrained  if we assume that the superpotential $W$ is defined for all values of $x$. 
There is no difficulty if we allow multi-valued superpotentials. In fact, the formula (\ref{Vnon-rel}) is nothing other than the `reduced' Hamilton-Jacobi equation for Hamilton's characteristic function $W$ and, as is well known, the characteristic function has branch points at turning points of the motion. Thus, {\it allowing for multi-valued superpotentials, a mechanical model has an $N=2$ pseudo-supersymmetric extension for every solution of its HJ equation}. 

\subsection{The BPS condition}

Given a solution of the HJ equation, and hence an $N=2$ pseudo-supersymmetric mechanics, we are faced with the problem of solving the constraints. One obvious way to do this is to set
\be\label{PSbarL}
P= \partial S \, , \qquad \bar\Lambda=0 \, .
\ee
Note that these equations are invariant under {\it both} pseudo-supersymmetries. This is obvious for $\bar{\bbQ}$ and true for $\bbQ$ because of cancellations in the Poisson bracket with $(P-\partial S)$.

Another way to solve the constraints  is to set
\be
K\left(P+\partial S\right) + 2J =0\, , \qquad \Lambda=0\, . 
\ee
If $K$ is non-invertible then this implies $v_mJ^m=0$ for $v$ in the kernel of $K$, which contradicts 
(\ref{restrictJ}), so this alternative is viable only if $K$ is invertible, and in this case it is equivalent to
\be\label{altBPS2}
\tilde P + K^{-1} J + \partial S =0\, ,  \qquad \Lambda=0\, , 
\ee
where
\be
\tilde P = P + K^{-1}J\, . 
\ee
In terms of $\tilde P$  the formula (\ref{facH}) becomes
\be
{\cal H} = \frac{1}{2}\left(\tilde P + K^{-1}J + \partial S\right)K\left(\tilde P -K^{-1}J -\partial S\right) 
\ee
from which we see that ${\cal H}$ is invariant under $\tilde P\to-\tilde P$. This means that one gets equivalent physics by taking $\tilde P\to -\tilde P$, but making this transformation in (\ref{altBPS2}) we recover the condition $P=\partial S$.  

Generically, {\it there are no other ways to solve the constraints}. For example, if one tries to consider a combination of the two alternatives just discussed, setting to zero  some of the components of $\bar\Lambda$ and the complementary components of $\Lambda$, then one finds that the Hamiltonian constraint is not solved because there remain $\Lambda\bar\Lambda$ terms. This may not happen for special choices of $S$ but then one expects additional symmetries, which plausibly render any additional possibilities equivalent to (\ref{PSbarL}). Given this, we conclude that  {\it the `first-order'  equations (\ref{foeqs}) are consequences of the pseudo-supersymmetry  constraints}. 

Although, the conditions (\ref{PSbarL}) preserve both pseudo-supersymmetries, a generic solution of the equations of motion for $(X,\Lambda)$ will break both of them.  To see this, first note that, 
\be\label{deltabarL}
\delta_{\bar\epsilon} \Lambda = 2\left(K\partial S +J\right) \bar\epsilon
\ee
for configurations satisfying (\ref{PSbarL}),  which shows that the $\bar{\bbQ}$ pseudo-supersymmetry will be broken unless $K\partial S + J=0$. This  condition cannot be met when $K$ is non-invertible, because of (\ref{restrictJ}), so in this case the $\bar{\bbQ}$ pseudo-supersymmetry is broken for {\it all} solutions of the equations of motion. Secondly, note that
\be
\delta_\epsilon X = -i \epsilon\Lambda\, , 
\ee
which shows that the $\bbQ$ pseudo-supersymmetry will be broken unless $\Lambda=0$. 

Our primary interest is in the original model, with $(X;P)$ phase space, and solutions of this model are found from solutions of the pseudo-supersymmetric model by settting 
\be\label{bosonic}
\Lambda=0\, , \qquad \bar\Lambda=0\, . 
\ee
Solutions of the equations of motion that have $\Lambda=\bar\Lambda=0$  initially will have $\Lambda=\bar\Lambda=0$ at all times, so  we may consistently restrict attention to configurations of this type. In this case, the condition for partial preservation of pseudo-supersymmetry is that the variations of $\Lambda$ and $\bar\Lambda$ vanish for non-zero  $\epsilon$ or non-zero $\bar\epsilon$.  The latter option is either not possible or equivalent to the former, so we assume that $\bar\epsilon=0$. The only non-zero pseudo-supersymmetry variation is then
\be
\delta_\epsilon\bar\Lambda = \left(P-\partial S\right)\epsilon\, . 
\ee
The condition for partial preservation of pseudo-supersymmetry is therefore  $P=\partial S$.  The `first-order' equation (\ref{foeqs}) of  the HJ formalism may thus be interpreted as a `BPS' condition for preservation of 1/2 pseudo-supersymmetry. Note, however, that this interpretation is simply a consequence of the constraints that define the model and the restriction to configurations 
satisfying (\ref{bosonic}): {\it all} solutions of the original mechanical model are BPS solutions of its pseudo-supersymmetric extension.

\subsection{Pseudo-Superspace}

Superfield methods may be used to make the pseudo-supersymmetries manifest. This is easily done for the $\bbQ$ pseudo-supersymmetry because  the corresponding `supercovariant derivative' $D$ satisfies $D^2\equiv 0$, which can be realized as $D=\partial/\partial\theta$ for independent real anticommuting variable $\theta$.  We now interpret $X$ and $\bar\Lambda$ as superfields with $\theta$-components\footnote{As is customary, we use the same symbol to denote a superfield and its lowest component. No confusion arises as long as one employs a formalism (as we do) that allows any component equation to be viewed as a superfield equaton.}
\be
DX^m = \Lambda^m \, , \qquad D\bar\Lambda_m = i\left(P_m -\partial_m S\right)\, . 
\ee
We also introduce the `fermion number'  superfield
\be
\bN =  \Lambda^m \bar\Lambda_m\, , 
\ee
which has the $\theta$-component
\be
D\bN =-i \bbQ\, . 
\ee
The other charge $\bar{\bbQ}$, now viewed as a superfield, has  $\theta$-component
\be\label{DQ}
D{\bar{\bbQ}} = 2i\bH\, . 
\ee
Now consider the Lagrangian
\be\label{BRSTexact}
\tilde L =-iD\left[ \dot X^m \bar\Lambda_m - \frac{1}{2}\ell\, \bar{\bbQ} +i \chi \, \bN\right]\, .
\ee
The Lagrange multipliers $\ell$ and $\chi$ are now superfields with $\theta$-components
\be
D\ell = 2\bar\chi\, , \qquad D\chi = iq\, , 
\ee
where $q$ is a new commuting variable. One finds that, on omitting a total derivative, that
\be
\tilde L = L +iq \bN \, , 
\ee
where $L$ is the Lagrangian of  (\ref{newlag}). The new variable $q$ therefore imposes a constraint of vanishing `fermion' number.  This constraint is satisfied automatically by the solution (\ref{PSbarL}) of the pseudo-supersymmetry constraints, which explains why we did not have to consider it previously. Also, it is consistent to set $q=0$, in which case we recover (\ref{newlag}) directly.

We have now shown how the $\bbQ$ pseudo-supersymmetry may be made manifest. To do the same for the $\bar{\bbQ}$ supersymmetry, the superfields $X^m$ and $\bar\Lambda_m$ of the above discussion would have to be combined into a single $N=2$ superfield depending on real anti-commuting variables $\theta$ and $\bar\theta$, but this will not be attempted here, in part because it is not obvious how to proceed when $K$ is non-invertible. 

\subsection{Quantum Theory}
\label{subsec:QT}

Let us now consider the quantum theory of the $N=2$ pseudo-supersymmetric extension of some model of mechanics. In contrast to the viewpoint adopted so far, the additional anti-commuting variables cannot be `set to zero'  at the end because they are now operators acting on Hilbert space. In particular, this means that  $\bbQ$, $\bar{\bbQ}$ and $\bH$ become operators (denoted by a `hat')  that span a superalgebra for which the only non-zero (anti)commutator is
\be\label{qqh}
\{ \hat{\bbQ}, \hat{\bar{\bbQ}}\} = 2\hat{\bH}\, . 
\ee
As a consequence, we have the operator identities
\be
{\hat{\bbQ}}^2 \equiv 0 \, , \qquad {\hat{\bar{\bbQ}}}^2 \equiv 0\, .  
\ee
For standard supersymmetry, $\hat{\bar{\bbQ}}$ is the hermitian adjoint of $\hat{\bbQ}$. For pseudo-supersymmetry,  $\hat{\bbQ}$ and $\hat{\bar{\bbQ}}$ are independent  hermitian operators, given 
that $\hat{\bH}$ is hermitian. However there are no nilpotent hermitian operators that act on a Hilbert space with positive definite norm. This might  have been anticipated from the non-unitarity of pseudo-supersymmetric field theories since one gets field equations from the first-quantization of particles. However, we set aside this difficulty  for the moment. 

Upon quantization, the Poisson bracket relations of (\ref{PBrelation}) become the (anti) commutation relations
\be
\left[\hat X^m,\hat P_n\right] = i \delta^m_n \, , \qquad \left\{\hat \Lambda^m,\hat{\bar\Lambda}_n\right\} = \delta^m_n\, , 
\ee
which we may realize by 
\be
\hat X^m =X^m \, ,\qquad  \hat\Lambda^m = \Lambda^m\, , \qquad
 \hat P_m = -i \partial_m \, , \qquad \hat{\bar\Lambda}_m = \partial/\partial\Lambda^m\, . 
\ee
It may be verified that the following operators are nilpotent:
\bea
\hat{\bbQ} &=& -i\Lambda^m\left(\partial_m -i\partial_m S\right)\, , \nonumber\\
\hat{\bar{\bbQ}} &=& -i\left[K^{mn}\left(\partial_n +i\partial_n S\right) + 2iJ^m - \Gamma^m_{pq}K^{qn}\, 
\frac{\partial}{\partial\Lambda^n} \Lambda^p\right]\frac{\partial}{\partial\Lambda^m}\, . 
\eea
Nilpotency of $\hat{\bbQ}$ is manifest and $\hat{\bar{\bbQ}}$ is nilpotent  provided that
\be
K^{pq}R^{[m}{}_{p\ell q}K^{n]\ell} =0\, , 
\ee
which, for invertible $K$, is the statement that the Ricci tensor is symmetric. The operator $\hat{\bH}$ may now be defined by (\ref{qqh}), and one finds that
\be
\hat{\bH}= \hat {\cal H} + \dots
\ee
where the dots indicate terms that are annihilated by $\partial/\partial\Lambda$, and
\be\label{calH}
\hat{\cal H} =  \frac{1}{2}\left[K^{mn}\left(\hat {\bP}_n + \partial_n S\right) + 2J^m \right] \left(\hat P_m -\partial_m S\right)\, .
\ee

The constraints are realized in the quantum theory by the following physical-state conditions on the wave-function $\Psi(X,\Lambda)$:
\be\label{Qpsi}
\hat{\bbQ} \Psi =0\, , \qquad  \hat{\bar{\bbQ}}\Psi =0\, . 
\ee
These are the only independent physical-state conditions since they imply that $\hat{\bH}\Psi=0$. We solve them along the same lines  as for the classical theory. Specifically, the classical conditions  (\ref{PSbarL}) become the following constraints on the wavefunction: 
\be
\partial_m \Psi = i\left(\partial_m S\right) \Psi\, , \qquad \partial \Psi /\partial\Lambda^m =0\, ,
\ee
for which the solution is 
\be\label{semiclass}
\Psi =  e^{iS(X)} \Psi_0
\ee
for constant $\Psi_0$.   This is a rather surprising result  because it implies that $\Psi$ has constant modulus; this is possible because $\hat{\cal H}$ is {\it not} hermitian, despite the (formal) hermiticity  of $\bH$. 

Another way of stating the above result is to observe that the effective action, after inclusion of quantum effects, equals the classical action, which means that the anti-commuting variables cancel out the quantum fluctuations of the original variables.  In effect, they allow a quantum description of a classical theory. This is what is also achieved by the  formalism of Gozzi et al. \cite{Gozzi:1986ge,Gozzi:1989bf} which involves additional anti-commuting variables that are interpreted as ghosts and anti-ghosts associated with a BRST and anti-BRST invariance. The phase superspace
in that  formalism has dimension $(4s|4s)$ rather than $(2s|2s)$, and there is no special role for the HJ equation, so it is unclear whether there is any connection to the formalism presented here. Nevertheless, the possibility of such a connection suggests that  the pseudo-supersymmetry charges introduced here should be interpreted as BRST charges. In this case the superspace expression (\ref{BRSTexact}) shows that the Lagrangian is BRST exact, and hence that we are dealing with a `topological' theory, as argued in  \cite{Gozzi:1989vv} for the formalism of  \cite{Gozzi:1986ge,Gozzi:1989bf}.

 \section{Relativistic Particle}
  \setcounter{equation}{0}

Let
\be\label{Qsrel}
\bbQ = \left(P_m - \partial_m S \right)\Lambda^m \, , \qquad
\bar {\bbQ} = \left(P_m + \partial_m S \right)\bar\Lambda^m\, , 
\ee
where $\bar\Lambda^m = \eta^{mn}\bar\Lambda_n$.  The only non-zero Poisson bracket  of these functions is that of (\ref{PBrelation}) provided that we 
choose
\be
\bH = \frac{1}{2}\eta^{mn} \left(P_m+\partial_m S\right) \left(P_n-\partial_n S\right)
-i \left(\partial_m\partial_n S\right) \Lambda^m\bar\Lambda^n\, . 
\ee
This corresponds to a potential given by 
\be\label{relU}
U=- \frac{1}{2} \left(\partial S\right)^2\, , 
\ee
which is just the HJ equation for our model. 

We may solve the pseudo-supersymmetry constraints in two equivalent ways:
\be
\!\!{\it either}\qquad P= \partial S\, , \quad \bar\Lambda=0\, , \qquad {\it or}\qquad
P=-\partial S\, , \quad \Lambda=0\, . 
\ee
For either choice we may restrict to solutions of the equations of motion for which both $\Lambda=0$ and $\bar\Lambda=0$, and in this case either the $\bbQ$ or the $\bar{\bbQ}$ pseudo-supersymmetry will be preserved, according to whether $P=\partial S$ or $P=-\partial S$.  The two possibilities are physically equivalent. Both pseudo-supersymmetries are preserved if (in addition) $P=0$, but this is the 2-momentum of the vacuum.

\subsection{Application to Cosmology}

The assumptions of homogeneity and isotropy in a model of gravity coupled to scalar fields with potential $V$, in $D$ spacetime dimensions, lead to an effective relativistic particle mechanics model with a Hamiltonian constraint of the form (\ref{HR}). In the conventions of \cite{Skenderis:2006jq},
the scale factor $a$ is written as  $\exp (\beta t)$, and the scalar potential of the effective particle mechanics model  is 
\be
U(t,x) = e^{2\alpha t} V(x) - \frac{k}{2\beta^2}e^{t/\alpha}\, , 
\ee
where $k=-1,0,1$ is the normalized curvature of spatial sections, and 
\be
\alpha = (D-1)\beta\, ,\qquad \beta = 1/\sqrt{2(D-1)(D-2)}\, . 
\ee
There are two cases in which the resulting  HJ equation may be solved by a separation of variables. The simplest is $V=0$, in which case 
\be
S= \pm \left(D-1\right) e^{t/(2\alpha)}\left[ e^{x/(2\alpha)} -k e^{-x/(2\alpha)}\right] \qquad (V=0). 
\ee

The other case for which the HJ equation may be solved by separation of variables is $k=0$, for which
\be
S= \pm 2e^{\alpha t} W(x)= 2 a^{D-1}W(x)  \qquad (k=0)\, , 
\ee
where the `comoving principal function' $W$ is $t$-independent and satisfies 
the `reduced HJ equation' \cite{Salopek:1990jq}
\be\label{Vcos}
V= -2\left[\left(W'\right)^2 - \alpha^2 W^2\right]  \, . \qquad \left[{\rm Cosmology}\right]\, . 
\ee
The equations (\ref{foeqs}) are now
\be\label{cosfo}
E= \mp 2\alpha e^{\alpha t} W(x)\, , \qquad p= \pm 2e^{\alpha t}W'(x)\, . 
\ee
Combining these with the equations of motion for $E$ and $p$ we arrive at the following first-order 
differential equations:
\be\label{focos}
\ell^{-1} \dot t = -2\alpha  e^{\alpha t} W(x)\, , \qquad \ell^{-1}\dot x = 2e^{\alpha t}W'(x)\, . 
\ee

We have focused on cosmology but exactly the same analysis applies to domain walls. The effective action is the same, except for a flip of the sign of the potential, so an application of HJ theory to flat domain walls \cite{de Boer:1999xf} leads again to the first-order equations (\ref{focos}) but with a `reduced HJ equation' of opposite sign for $V$: 
\be\label{VDW}
V= 2\left[\left(W'\right)^2 - \alpha^2 W^2\right]  \, . \qquad \left[{\rm Domain \ Wall}\right]
\ee
Remarkably, this formula coincides with the `supergravity-inspired' formula for $V$ introduced in \cite{Boucher:1984yx,Townsend:1984iu}; in fact, for $D=3$ it {\it is} the supergravity formula for $V$ in terms of a superpotential $W$.  In the context of supergravity domain walls, first-order equations consistent with (\ref{focos}) arise as `BPS-type' conditions for the preservation of 1/2 supersymmetry  \cite{Cvetic:1992bf,Cvetic:1996vr}, and they can be found precisely in the HJ form (\ref{focos})  by `supergravity-inspired' methods \cite{Skenderis:1999mm,Freedman:1999gp}. Specifically, they arise as integrability conditions for the existence of Killing spinors defined in the context of  `fake supergravity' \cite{Freedman:2003ax,Sonner:2005sj}. It was pointed out in \cite{Skenderis:2006jq} that this `coincidence' between the HJ and fake supergravity approaches to domain walls also applies to cosmology but the (fake) Killing spinors become (fake) pseudo-Killing spinors, defined by an analytic continuation for which $W\to iW$.  As mentioned in the introduction,  it has been shown recently  that  fake pseudo-Killing spinors  may be `genuine' Killing spinors of a `pseudo-supergravity' theory, defined by analytic continuation of a standard supergravity theory to one with spinors obeying   `twisted reality' conditions. 

\subsubsection{Non-flat cosmology}

It was shown in \cite{Skenderis:2006jq} that non-flat cosmologies are determined by the following first-order equations involving a complex superpotential $Z$: 
\be\label{Zpot}
E= \mp 2\alpha e^{\alpha t} \, \left[\frac{
{\cal R}e\left(\bar Z Z'\right)}{|Z'|}\right]\, , \qquad
p= \pm e^{\alpha t}\, |Z'|\, . 
\ee
These equation reduce to those of (\ref{cosfo}) when $Z=W$, which applies when $k=0$. One may ask what the relation of these equations is to the first-order equations of HJ theory, 
\be\label{HJJ}
E= -\partial_t S \, , \qquad p= \partial_x S\, . 
\ee
If we try to combine (\ref{HJJ}) with (\ref{Zpot}), we arrive at the equation
\be\label{dS}
\pm dS = \omega \equiv 2\alpha e^{\alpha t} \left[ \frac{{\cal R}e\, \left(\bar Z Z'\right)}{\left|Z'\right|}\right] dt
+ e^{\alpha t} \left|Z'\right| dx\, .
\ee
However, the 1-form $\omega$ is not closed for $k\ne0$; using eq, (4.16) of \cite{Skenderis:2006rr}, one finds that
\be
d\omega = \frac{k}{2\beta}\left|Z'\right|^{-1} e^{\left(\alpha-2\beta\right)t} dt\wedge dx\, .
\ee
It follows that (\ref{dS}) cannot be integrated to give a function $S(t,x)$. An expression for $S$ for $k\ne0$ was found in \cite{Skenderis:2006rr}, but the construction assumed that one is given a solution of the equations of motion, in which case the variables $(t,x)$ are not independent.

For $k=1$ cosmologies, the `first-order equations that follow from (\ref{Zpot}) on using the equations of motion for $E$ and $p$ are integrability conditions for the existence of pseudo-Killing spinors, and in this `field theoretic' sense they may be interpreted as `BPS' conditions\footnote{This property distinguishes these `first-order equations from others proposed previously \cite{Bazeia:2005tj}.}, as for $k=0$.  Again,  there is a parallel story for domain walls, most of which came first; we focus here on cosmology for convenience of presentation, but there is one point that is easier to understand from the domain-wall perspective: we do not expect to find Killing spinors for `de Sitter sliced' walls, and this corresponds to the statement that we should not expect to find pseudo-Killing spinors for $k=-1$
cosmologies. This is indeed the case, even though there are first order equations for any $k$
This state of affairs should be contrasted with the particle mechanics BPS interpretation of the 
equations (\ref{HJJ}), which applies {\it for all} $k$. Moreover, in the particle mechanics BPS interpretation, the superpotential is real {\it for all} $k$, being identified with the `comoving 
principal function' $W$.  

These comments  suggest  that there is no simple general connection between the field theoretic BPS  interpretation of the first-order equations for cosmology that arise from the existence of pseudo-Killing spinors and the `intrinsic' BPS interpretation proposed here in the context of an $N=2$ pseudo-supersymmetric extension of mechanics.

\section{Non-relativistic particle}

For a non-relativistic particle in a potential $U(t,x)$,  the pseudo-supersymmetric extension is found by choosing\footnote{This is suggested by a `worldline supergravity' Lagrangian proposed 
in  \cite{Gomis:1985uf}.}

\be
\bbQ =  \left(p - \frac{\partial S}{\partial x} \right)\lambda +
 \left(E+ \frac{\partial S}{\partial t}\right)\psi \, , \qquad
 \bar {\bbQ} = \left(p + \frac{\partial S}{\partial x} \right)\bar\lambda + 2\bar\psi\, ,  
\ee
where $S(t,x)$ is a superpotential. The only non-zero Poisson bracket  is that of (\ref{PBrelation}) provided that we also choose
\be
\bH =- E  + \frac{1}{2}p^2  + U  -i \left(\frac{\partial^2 S}{\partial x^2}\right)\lambda\bar\lambda + 
i\left(\frac{\partial^2 S}{\partial x\partial t}\right)\psi\bar\lambda\, ,
\ee
where
\be\label{bV}
U= - \frac{\partial S}{\partial t} - \frac{1}{2}\left(\frac{\partial S}{\partial x}\right)^2\, ,  
\ee
which is the HJ equation of our model. 

To satisfy the constraints we set 
\be
E= -\frac{\partial S}{\partial t}\, , \qquad p= \frac{\partial S}{\partial x}\, ,  
\ee
and
\be
\bar\psi =0 \, , \qquad \bar\lambda =0\, . 
\ee
From the infinitesimal pseudo-supersymmetry transformations
\bea
\delta\psi &=& -2\bar\epsilon\, , \qquad
\delta \lambda = \left(p+ \frac{\partial S}{\partial x}\right)\bar\epsilon \nonumber\\
\delta\bar\psi &=& -\left(E+\frac{\partial S}{\partial t}\right)\epsilon\, , \qquad 
\delta\bar\lambda = \left(p-\frac{\partial S}{\partial x}\right)\epsilon\, , 
\eea
we see that the $\bar{\bbQ}$ pseudo-supersymmetry is necessarily broken. Provided that $\psi=\lambda=0$, the $\bbQ$ pseudo-supersymmetry is preserved.

\subsection{Time-independent potential}

It will prove instructive to analyse in more detail the special case for which 
\be\label{character}
S(t,x) = W(x) -E_0\,t\, , 
\ee
for constant $E_0$, and a function $W$ that  can be interpreted as Hamilton's ``characteristic'' function. 
In this case we have the $t$-independent  Hamiltonian constraint function
\be\label{H1}
\bH = -E +  \frac{1}{2} p^2 +V   +i W'{}' \bar\lambda \lambda\, ,   
\ee
where 
\be\label{H2}
V= E_0  -\frac{1}{2} \left(W'\right)^2\, ,  
\ee
which is the `reduced' HJ equation appropriate for a principal function of the assumed form. 
The pseudo-supersymmetry constraint functions are now
\be
\bbQ =  \left(p - W'\right)\lambda +
 \left(E-E_0\right)\psi \, , \qquad
 \bar {\bbQ} = \left(p + W'\right)\bar\lambda + 2\bar\psi\, ,  
\ee
and they generate the following infinitesimal pseudo-supersymmetry transformations:
\bea\label{canontrans}
\delta x &=& -i\left(\epsilon\lambda + \bar\epsilon\bar\lambda\right) \, , \qquad
\delta p = -i\left(\epsilon\lambda - \bar\epsilon\bar\lambda\right)W'{}' \nonumber\\
\delta t &=& i\epsilon\psi \, , \qquad \delta E=0 \nonumber\\
\delta \lambda &=& \left(p+W'\right)\bar\epsilon\, , \qquad \delta\bar\lambda = \left(p-W'\right)\epsilon \nonumber\\
\delta\psi &=& -2\bar\epsilon\, , \qquad \delta\bar\psi = -\left(E-E_0\right)\epsilon\, . 
\eea
The Lagrangian (\ref{newlag}) transforms into a total derivative under these transformations provided the  Lagrange multipliers are  assigned the transformations
\be
\delta\chi = \dot \epsilon\, , \qquad \delta\bar\chi = \dot{\bar\chi}\, , \qquad \delta\ell = -2i\left(\epsilon\bar\chi + \bar\epsilon\chi\right)\, . 
\ee

To simplify things we will now partially fix the local pseudo-supersymmetry transformations by the gauge choice \be
\chi=0\, , \qquad \bar\chi=0\, . 
\ee
This leaves a residual invariance under the transformations of (\ref{canontrans}) for {\it constant} $(\epsilon, \bar\epsilon)$. We must therefore still impose the constraints $\bbQ=0$ and $\bar{\bbQ}= 0$ at some initial time, but they will then hold at all times as a consequence of the equations of motion. The advantage of this gauge choice is that  we now have the much simpler Lagrangian. 
\be
L= \dot x p + i\bar\lambda\dot\lambda -\ell\left[E_0 + \frac{1}{2} p^2 -\frac{1}{2}\left(W'\right)^2 
+iW'{}' \bar\lambda \lambda\right]  -i\bar\psi\dot\psi + E\left(\ell-\dot t\right)\, . 
\ee
We may further simplify by using the equations of motion of the conjugate pair
$(\psi,\bar\psi)$, which imply that
\be
\psi=\psi_0 \, , \qquad \bar\psi = \bar\psi_0\, , 
\ee
for anti-commuting constants $(\psi_0,\bar\psi_0)$, together with the equations of motion 
of the conjugate pairs $(t,-E)$:
\be
\dot E=0\, , \qquad \ell=\dot t\, . 
\ee
Ultimately, one finds that the constant value of $E$ must be $E_0$ and we shall now assume this in order to shorten the presentation.  We now have the even simpler Lagrangian 
\be\label{evenS}
L= \dot x p + i\bar\lambda\dot\lambda -\dot t \, \bH\, , 
\ee
where, from (\ref{H1}) and (\ref{H2}), 
\be\label{simpH}
\bH =  \frac{1}{2}p^2 - \frac{1}{2}\left(W'\right)^2 +iW'{}'\bar\lambda\lambda\, . 
\ee
The pseudo-supersymmetry Noether charges are now
\be
\bbQ =  \left(p - W'\right)\lambda \, , \qquad
 \bar {\bbQ} = \left(p + W'\right)\bar\lambda + 2\bar\psi_0\, ,  
\ee
and they generate the transformations\footnote{The transformation $\delta t= i\epsilon\psi_0$  is not generated by these simplified Noether charges because of the simplification we made in eliminating the variable $E$,  but this is irrelevant because there is a manifest  {\it independent}  invariance under constant shifts of $t$ that we may combine with the pseudo-supersymmetry to  arrange for $t$ to be inert.}
\bea\label{simple1}
\delta x &=& -i\left(\epsilon\lambda + \bar\epsilon\bar\lambda\right) \, , \qquad
\delta p = -i\left(\epsilon\lambda - \bar\epsilon\bar\lambda\right)W'{}' \, , \nonumber\\
\delta \lambda &=& \left(p+W'\right)\bar\epsilon\, , \qquad \delta\bar\lambda = \left(p-W'\right)\epsilon \, . 
\eea
together with 
\be\label{simple2}
\delta\psi_0 = -2\bar\epsilon\, . 
\ee

Finally, we may fix the time-reparametrization invariance by the gauge choice
\be
\dot t=1\, 
\ee
to arrive at the Lagrangian
\be\label{gaugefixed}
L= \dot x p + i\bar\lambda\dot\lambda - \bH \, ,
\ee
where $\bH$, which is still given by (\ref{simpH}), may now be interpreted as the Hamiltonian.  
The gauge choice is again only partial because it leaves a residual invariance under rigid time translations, and this means that we must set $\bH=0$ at some initial time, now as a zero-charge condition which holds at all times as a consequence of the equations of motion.  
The zero charge conditions may be solved in the same way that we previously solved the phase superspace constraints. We must arrange for  $\bbQ=0$ and $\bar{\bbQ}=0$ in such a way that one of the two pseudo-supersymmetries is preserved when all anti-commuting variables are zero. 

Although we could omit the transformation (\ref{simple2}) on the grounds that this is an independent symmetry (trivially since the Lagrangian is independent of $\psi_0$)  we will retain it for the moment as 
a relic of the asymmetry between $\bbQ$ and $\bar{\bbQ}$ that is inherent in the physics of the non-relativistic particle: this asymmetry is due to a choice of positive rather than negative energy in taking the non-relativistic limit. In this case, the  symmetry generated by $\bar{\bbQ}$ is necessarily broken, so we must preserve the symmetry generated by $\bbQ$, and this requires
\be
\qquad p= W'\, , \qquad \lambda=0\, . 
\ee
We see that the equation relating the momentum to Hamilton's characteristic function is a BPS condition
for 1/2 pseudo-supersymmetry. This should be combined with the equation (\ref{H2}) for $V$ in terms of the superpotential $W$, which we
may rewrite as
\be\label{Wprime}
W'(x) = \pm \sqrt{2\left(E_0-V(x)\right)}
\ee
and interpret as the reduced HJ equation for Hamilton's characteristic function $W$ in terms of $V$. 
There is a solution only for $E_0>V$ so if $V$ is unbounded from above there is no choice of $E_0$ for which $W(x)$ is defined for all $x$. This is of course a well known result in mechanics. The corresponding result for superpotentials, and its implications in the context of supersymmetric domain walls was recently discussed in \cite{Sonner:2007cp}.

\subsubsection{Relation to supersymmetric mechanics}

If we  eliminate $p$ from the Lagrangian (\ref{gaugefixed}) by using its equation of motion $p=\dot x$, then 
we arrive at the equivalent Lagrangian
\be\label{PSQM}
L= \frac{1}{2} \dot x^2   + i\bar\lambda\dot\lambda+ \frac{1}{2}\left(W'\right)^2 
+ iW'{}' \lambda\bar\lambda\, , 
\ee
which is invariant under the $N=2$ pseudo-supersymmetry transformations
\be\label{analytic}
\delta x =-i\left(\epsilon\lambda + \bar\epsilon\bar\lambda\right)\, , \qquad 
\delta\lambda = \left(\dot x+W'\right)\bar\epsilon\, , \qquad 
\delta\bar\lambda = \left(\dot x-W'\right)\epsilon\, . 
\ee
This is similar to a model of $N=2$ supersymmetric mechanics, but with `wrong sign' potential, In fact, it is related to $N=2$ supersymmetric mechanics by complex analytic continuation, as we now explain. 

Complexify all the dynamical variables $(x,\lambda,\bar\lambda)$. The result is a complex Lagrangian, but  one that depends analytically on them, and  it remains invariant under the transformations (\ref{analytic}) with complixified $(\epsilon,\bar\epsilon)$, which are now analytic transformations. Clearly, we recover the original Lagrangian on restricting all fields to be real, but we may now ask whether there is some other way to get a real Lagrangian\footnote{Here we follow the logic presented in \cite{Bergshoeff:2007cg}  for supergravity.}. If there is, it must involve an analytic continuation of the real variables of  (\ref{PSQM}). Consider the analytic continuation that effects
\be
W\to iW\, . 
\ee 
This yields the new Lagrangian
\be\label{SQM}
L= \frac{1}{2} \dot x^2 + i\bar\lambda\dot\lambda - \frac{1}{2}\left(W'\right)^2 
- W'{}' \lambda\bar\lambda\, . 
\ee
This is again real provided we choose $\lambda$ to be complex with complex conjugate $\bar\lambda$.  
The same analytical continuation of the transformations (\ref{analytic}) yields
\be\label{susy}
\delta x =-i\left(\epsilon\lambda + \bar\epsilon\bar\lambda\right)\, , \qquad 
\delta\lambda = \left(\dot x+ iW'\right)\bar\epsilon\, , \qquad 
\delta\bar\lambda = \left(\dot x-iW'\right)\epsilon\, . 
\ee
where $\epsilon$ is now a complex parameter with complex conjugate $\bar\epsilon$. This is now a standard $N=2$ supersymmetric mechanics model with superpotential $W$.

\section{Discussion}

We have derived the Hamilton-Jacobi formulation of  mechanics by considering the possible $N=2$ locally pseudo-supersymmetric extensions of reparametrization invariant Lagrangians for mechanical systems for which the Hamiltonian is  no more than quadratic in momenta, as can usually be arranged. The Hamilton-Jacobi equation arises as a condition for pseudo-supersymmetrizability, with the superpotential taking the role of Hamilton's principal function; one need not worry about the fact that there may not always be a solution because the superpotential is allowed to be multi-valued, with branch points, and  there is no superpotential only when there is no real solution of  the Hamilton-Jacobi equation. 

The `first-order'  equations of the HJ formalism relating the momenta to Hamilton's principal function arise as solutions to the constraints associated to the local symmetries, and if one sets all anti-commuting variables to zero then these constraints may be interpreted as BPS conditions arising from the preservation of 1/2 pseudo-supersymmetry. This is very likely related to a `field theoretic' BPS interpretation that arises from consideration of pseudo-Killing spinors in cosmological spacetimes for flat universes, which can be realized in some cases as supersymmetric solutions of pseudo-supergravity theories. However, a comparison with results of this `field theoretic'  BPS interpretation  for non-flat cosmologies makes it appear unlikely that there is any simple general relation. One possible 
route to a further investigation of this point would be to extend the considerations of this paper to mechanical models with an infinite number of degrees of freedom. After all, a field theory can be 
viewed as a model of mechanics on an infinite-dimesional space. From this perspective, the results
obtained here should also apply to field theory,  although functions such as Hamilton's principal 
function then become functionals. 

It was stated in the introduction that the pseudo-supersymmetric formulation of HJ theory makes it 
clear why the HJ equation is equivalent, if taken together with the `first-order' equations that define
the momenta, to the Hamilton equations of motion. The reason is that in a reparametrization-invariant formalism, as used here, all the dynamics is encoded in the constraints. Thus, the dynamics of the model defined by (\ref{startlag})  is encoded in the Hamiltonian constraint function ${\cal H}$. However, to get to the HJ equation we had to use the relations (\ref{foeqs}) for the momenta in terms of the principal function, and this needs a separate motivation (e.g. involving considerations of canonical transformations). In contrast, these relations are implied by the constraints in the pseudo-supersymmetrized model, while the model itself is determined by a solution of the HJ equations.  
Solutions of the original model are then found as solutions of the pseudo-supersymmetry constraints 
with vanishing anti-commuting variables,  all of which preserve (at least) 1/2 of the pseudo-supersymmetry.  We thus have a correspondence between solutions of the Hamilton equations (\ref{HEM})  and solutions obtained by  integration of the `BPS' equations, alias the `first-order'  equations (\ref{foeqs}), for a given a solution of the HJ equation.

Classical mechanics has been the focus of this work but it is natural to ask whether the anti-commuting variables arising from pseudo-supersymmetry have a role to play in the quantum theory. On the one hand, this seems to ruled out by the impossibility of a realization of the quantum pseudo-symmetry algebra via operators acting on a Hilbert space with positive definite norm. On the other hand, a simple computation of the wavefunction shows that the effect of the anti-commuting variables is to cancel the quantum fluctuations of the original variables, so the model is effectively still classical. This suggests
a possible BRST interpretation of the pseudo-supersymmetry algebra, as in the  formalism of Gozzi et al. \cite{Gozzi:1986ge,Gozzi:1989bf}, which has some features in common with the formalism presented here. It would be satisfying if the two formalisms could be related;  in any case, a better understanding of 
the quantum theory  is clearly desirable. In view of the many  insights into quantum mechanics provided by  supersymmetric quantum mechanics (see e.g. \cite{Cooper:2001zd}) it seems reasonable to  hope  that quantum pseudo-supersymmetric mechanics will be useful too.

\subsection*{Acknowledgments}
I am grateful to Gary Gibbons, Joaquim Gomis and Kostas Skenderis for helpful discussions on Hamilton-Jacobi theory, and for bringing the work of Gozzi et al. to my attention. This work was supported by an EPSRC Senior Research Fellowship.

\end{document}